\documentclass{article}
\usepackage{graphicx} % Required for inserting images
\usepackage{graphics}
\usepackage{soul}
\usepackage{cite}
\usepackage{amsmath,amssymb}
\usepackage{xcolor}
\usepackage{float}
\usepackage{soul}
\DeclareUnicodeCharacter{2061}{}

\title{Quantum Particle Creation by Cosmic Strings in de Sitter Spacetime}
\author{
    Bilgehan Barış \"{O}NER  and \"{O}zlem YEŞİLTAŞ
}
\date{
    \begin{tabular}{c c}
       \small  Gazi University, Faculty of Science, Physics Department, 06500, Teknikokullar, Ankara, TURKEY \\
            \end{tabular} \\[10pt]
    March 2025
}
\begin{document}

\maketitle

\begin{abstract}

This paper explores the phenomenon of particle creation associated with cosmic strings in de Sitter spacetime, a model that represents the universe's exponential expansion. We examine how the presence of cosmic strings in a de Sitter background affects particle production, focusing on the roles of string tension and angular deficits. Utilizing the Klein-Gordon equation adapted to curved spacetime with cosmic string defects, we derive solutions expressed through hypergeometric functions to describe particle states. Our findings highlight how string properties influence particle creation rates and energy distributions. By analyzing both point-like and linear potentials near the string, we determine exact solutions, investigate asymptotic behaviors, and calculate particle creation probabilities using Bogoliubov transformations.
\end{abstract}
\section{Introduction}

Cosmic strings (CS) have long captured the attention of researchers due to their fascinating structure and profound influence on the surrounding spacetime \cite{Kibble}. Tom W. B. Kibble was the first to propose the idea of cosmic strings in his seminal paper titled "Topology of Cosmic Domains and Strings" in 1976. In Kibble's model, cosmic strings are one-dimensional objects that form when a gauge field associated with a broken symmetry condenses, creating narrow, dense regions of energy in space. These strings stretch across vast distances in the universe and can potentially influence its structure and evolution. These one-dimensional theoretical objects hold great potential for advancing our understanding of the universe and its underlying physical laws. CS are distinguished by their extremely dense spatial energy distributions, resulting in strong gravitational effects such as gravitational lensing \cite{Vilenkin}, the emission of gravitational waves \cite{Damour1}, signatures in the cosmic microwave background \cite{Hindmarsh}, particle deflection \cite{Pogosian}, and the formation of cusps and kinks \cite{Blanco-Pillado1}, among others.

De Sitter spacetime, characterized by a positive cosmological constant, describes a universe undergoing exponential expansion. Studying CS within de Sitter spacetime allows for a more realistic model of particle interactions, incorporating the accelerated expansion of the universe into the calculations \cite{Mello}, \cite{Bezerra}, \cite{sucu}. One of the most intriguing and important effects related to CS is particle production \cite{Dias}, \cite{Belbaki}. This phenomenon occurs when CS, due to their strong gravitational fields and oscillatory motion, induce the creation of particle-antiparticle pairs. These pairs are produced near the horizon of the gravitational defect formed by the string. The ability of cosmic strings to create particles stems from quantum field theory in curved spacetime, where fluctuations in the field or string oscillations can lead to particle creation from the vacuum.

The study of particle production is crucial for several reasons, one of them is observational signatures \cite{Srednicki}: The radiation emitted by particle production processes could serve as a detectable signature of CS. This radiation is analogous to Hawking radiation from black holes, and it offers a potential indirect method of detecting these elusive objects. Also, particle production can act as a mechanism through which CS lose energy \cite{Damour2}. As cosmic strings oscillate and stretch, they radiate energy in the form of particles, which could affect their dynamics and evolution in the universe. Understanding particle production is essential for modeling the evolution of CS networks in the early universe. These calculations help in predicting how CS interact with surrounding matter and radiation fields, influencing cosmic structure formation and potentially leaving imprints on observable quantities, such as the cosmic microwave background (CMB) \cite{Albrecht}. As a link to fundamental physics; particle production by CS provides insight into the interactions between fields and spacetime \cite{Polchinski}. Studying these processes can reveal new aspects of quantum field theory, gravitational physics, and even aspects of string theory or higher-dimensional theories where cosmic strings are prominent. For instance, in the case of particle-antiparticle pair creation near a cosmic string, if one particle gains sufficient energy, it escapes the gravitational influence of the string and becomes observable in the universe, while the other particle falls into the defect, potentially carrying negative energy. This process mirrors Hawking radiation around black holes, suggesting that cosmic strings could exhibit a similar emission mechanism. The emitted particles could be detected as cosmic radiation, providing direct evidence of the existence of CS. More recent studies can be found in \cite{Gorghetto,Blanco-Pillado2,Bellucci}.

Therefore, particle production calculations are not only important for theoretical insights but also for identifying potential observational signatures of cosmic strings, which could confirm their existence and role in the evolution of the universe.

All aforementioned topics show that the impacts of cosmic strings must be examined in an expanding universe since they may have created during inflation. Particle production is altered by de Sitter space, and cosmic strings may intensify this effect. Comprehending cosmic string interactions in de Sitter offers a novel test for quantum field theory in curved spacetime, which has consequences for gravitation and astroparticle physics. It may also bring new insights into the early cosmos and gravitational waves.

In this article, we investigate the effects of cosmic string tension and angular deficit on the properties of particle creation within de Sitter spacetime. In Section II, we begin by reformulating the Klein-Gordon equation in the curved geometry of de Sitter spacetime with a cosmic string, introducing the necessary transformations to facilitate the analysis. Then,  we derive and analyze radial solutions under different scenarios, initially, including the presence of a point-like source. In Section III, we continue with  a dense cosmic string, and a charge-free medium. We also extend our investigation to the case of a linear potential, exploring how these configurations modify particle creation probabilities through Bogoliubov transformations. Finally, we discuss the implications of our findings for the symmetry of particle creation and annihilation, the unitarity of quantum processes, and potential observational signatures, such as scalar radiation and cosmic string dynamics in the early universe.

\section{Reformatting the Klein-Gordon Equation}
Dynamic de Sitter space is an expanding space-time model of cosmos that is frequently investigated in the context of general relativity and cosmology. This metric allows the study of cosmic string effects in an expanding universe, which is crucial for understanding early-universe phenomena. Time-dependent scale factors reflecting the expansion of the universe would be incorporated into the metric which is usually considered as exponential. In a region under the influence of the cosmic string, the geometry of dynamic de Sitter space can be more complicated than that of the standard. Since cosmic strings can be modeled as one-dimensional topological defects, it is useful to express the axial symmetry within cylindrical coordinates while dealing with these strings. To do this let us begin with a cosmic string line element in de Sitter space

\begin{equation}\label{ds2}
ds^2 = d{t_c}^2 - e^{2 H t_c} ( {dr_c}^2  +  \alpha^2 {r_c}^2 d\varphi_c^2  +  dz_c^2).
\end{equation}
where the subindex $c$ is for cylindrical coordinates, $r \geq 0, -\infty < z < \infty$. We choose the Minkowksi metric signature of [+,-,-,-] and here $H$ is the Hubble Constant which equals $\sqrt{\Lambda / 3}$, $\Lambda$ denotes the positive cosmological constant. Here, the string tension directly affects the geometry of the spacetime around a cosmic string, leading to an angular deficit $2 \pi (1 - \alpha)$ where $\alpha=1-4G\mu$, $\mu$ is the string tension, $G$ is the gravitational constant. Let us implement a similar transformation given by \cite{Mello} for a more convenient mathematical analysis in a static space. To move from the dynamic cylindrical coordinates in de Sitter spacetime to static spherical one, we use the following transformations \cite{Mello}:

\begin{equation}\label{donusum}
  t_c=-t+\frac{1}{2H} \ln (1-H^{2}r^{2}),~~ r_c=\frac{r e^{H t}\sin \theta}{\sqrt{1-H^{2}r^{2}}}, ~~z_c=\frac{r e^{H t}\cos \theta}{\sqrt{1-H^{2}r^{2}}},~~\varphi_c=\varphi.
\end{equation}

Here, while the radial and time components are scaled by the exponential expansion, the angular part remains tied to the intrinsic conical geometry imposed by the cosmic string. This transformation vanishes the explicit time dependency of the metric and confines the spatial distance within $0 \leq r < H^{-1}$ Even though the transformation is mathematically convenient an inverse transformation of $r^2=(r_c^2+z_c^2 ) e^{2Ht_c}$ should be implemented to make a better interpretation of the results. Then, we reach the covariant metric tensor $g_{\mu \nu}$ near the string as below:

\begin{equation}\label{metrikTensor}
g = \begin{pmatrix}
1-H^2 r^2 & 0 & 0 & 0 \\
0 & \frac{-1}{1-H^2 r^2} & 0 & 0 \\
0 & 0 & -r^2  & 0 \\
0 & 0 & 0 & -\alpha^2 r^2 sin^2 \theta
\end{pmatrix}
\end{equation}

\noindent where we denote statical spherical coordinates as $x=(t,r,\theta,\varphi)$. The Klein-Gordon Equation (KG) for a scalar field in a curved spacetime is

\begin{equation}\label{KG}
\left[ -\frac{1}{\sqrt{-g}} D_\mu g^{\mu\nu} \sqrt{-g} D_\nu + M^2 \right] \Phi = 0,
\end{equation}

\noindent where $g$ is determinant of the covariant metric tensor and $g^{\mu\nu}$ is for the contravariant metric tensor. The Klein-Gordon Equation may be rewritten as below assuming a four-vector potential of $A=(A_0(r),0,0,0)$

\begin{equation} \label{KGFull}
\resizebox{\textwidth}{!}{%
$\begin{aligned}
\bigg[
\frac{-{({\partial_t}+iqA_{0}(r))^2}}{1 - H^2 r^2}  +
\frac{\big[\partial_r \big(r^2 (1-H^2r^2) \partial_r \big) \big]}{r^2}  +
\frac{(cot\theta \, \partial_\theta + {{\partial_\theta}^2})}{r^2}  +
\frac{{\partial_\varphi}^2}{\alpha^2 r^2 sin^2\theta}  + M^2
\bigg] \Phi =0.
\end{aligned}$
}
\end{equation}

Here, $M$ denotes the mass of the particle. Assuming a wavefunction as given below:
\begin{equation}\label{wf}
\Phi = \phi(r) F(\theta) {e}^{-i(\epsilon t - m \varphi)},
\end{equation}
and applying the method of separation of variables to the Klein-Gordon equation, the angular part of the equation can be easily isolated. This angular part has already been provided in \cite{Belbaki}, where it is mentioned that the corresponding solution involves generalized Legendre functions. The angular part of the Klein-Gordon equation, after separation of variables in a curved spacetime with a cosmic string, can be expressed as follows:

\begin{equation} \label{KGAngular}
\bigg[ \frac{1}{sin\theta} \frac{d}{d \theta} (sin\theta \frac{d}{d \theta} ) - \frac{m^2}{\alpha^2 sin^2\theta} \bigg] F(\theta) = -l_{\alpha}  (l_{\alpha}  +1) F(\theta)
\end{equation}

\noindent where $l_{\alpha} = n + \frac{|m|}{\alpha} $ such that $n$ is a non-negative integer  and $m$ is the azimuthal quantum number associated with rotation around the $z$-axis. In a similar way, the radial equation may be separated. In this context the rest of the paper analyzes three different cases. Initially, radial solutions are derived for the Klein-Gordon equation under the influence of a point-like source in a static spacetime, with the vector potential $A_0(r)=Q/r$ representing the most general case. Subsequently, a similar approach is applied to a scenario involving a dense cosmic string, contrasting it with a charge-free medium where $A=0$. Finally, the analysis is extended to a linear potential case, where the potential $V(r) \propto r$  models the influence of an external field within a localized spatial region, capturing the effects of a cosmic string immersed in a uniform background field.

\subsection{The Exact Solution for Point-like Source} \label{Exact_pointlike}

The potential of a point source with a charge $Q$ in de Sitter spacetime is expressed as  below when $r \ll 1/H$
\begin{equation}\label{vr}
  V(r)=  \frac{Q}{r}
\end{equation}

\noindent derived as a solution to the generalized Poisson's Equation in curved spacetime (please see Appendix A). For simplicity, the electromagnetic four-potential can therefore be approximated as:

\begin{equation} \label{A}
 \bold{A}=\bigg[\frac{Q}{r} + C_0,0,0,0 \bigg]
\end{equation}

\noindent where $C_0$ is a constant that represents a shift in the potential and the radial part of the wave equation becomes

\begin{equation} \label{KGQ_general}
\resizebox{\textwidth}{!}{%
$\begin{aligned}
(1-H^2r^2) \phi(r)^{''} + \frac{2}{r} (1-2H^2r^2) \phi(r)^{'} +
\bigg(\frac{ (\epsilon - q \frac{Q}{r} - q C_0)^2  }{1-H^2r^2} + M^2 - \frac{l_{\alpha} (l_{\alpha}+1)}{r^2}  \bigg) \phi(r) = 0.
\end{aligned}$
}
\end{equation}

The solutions of this equation are given by generalized Heun functions, $\mathcal{H_G}$, such that

\begin{equation}
\resizebox{0.9\textwidth}{!}{%
$\begin{matrix}
\phi_1 (r) = f_{1}(r) \mathcal{H_G} \bigg(-1, \frac{(E_Q-1)  (1 + L_Q)}{-2} + T_1 , \frac{H (E_Q + L_Q + 1) + \mu}{2 H}, \frac{H (E_Q + L_Q + 1) \mu - (\Tilde{M}^2 + \Tilde{\epsilon}^2)}{2 \mu H}, 1 + L_Q, E_Q, -Hr \bigg) \\
\phi_2 (r) = f_{2}(r) \mathcal{H_G} \bigg(-1, \frac{(E_Q-1)  (1 - L_Q)}{-2} + T_2, \frac{H (E_Q - L_Q + 1) + \mu}{2 H}, \frac{H (E_Q - L_Q + 1) \mu - (\Tilde{M}^2 + \Tilde{\epsilon}^2)}{2 \mu H}, 1-L_Q, E_Q, -Hr \bigg) \label{KGQ_general_solutions}

\end{matrix}$
}
\end{equation}

\noindent or any of their linear combinations where

\begin{equation}
\begin{matrix}
f_{1}(r) = (-1)^{\frac{E_Q}{2}} r^{\frac{-1 + L_Q}{2}} (1+ Hr)^{ \frac{E_Q - 1 }{2}} (Hr - 1)^{\frac{- \Tilde{\epsilon}}{2H}} \\
f_{2}(r) = (-1)^{\frac{E_Q}{2}} r^{\frac{-1 - L_Q}{2}} (1+ Hr)^{ \frac{E_Q - 1 }{2}} (Hr - 1)^{\frac{- \Tilde{\epsilon}}{2H}} \\ \\
T_1 =\frac{4Qq (\epsilon - q C_0) \mu + i (1 + L_Q) \Tilde{\epsilon} \, \mu^{*}}{ 2 \mu H}, \quad
T_2 =  \frac{4Qq (\epsilon - q C_0) \mu + i (1 - L_Q) \Tilde{\epsilon} \, \mu^{*}}{ 2 \mu H} \\
E_Q = 1 + i(Q q + \frac{(\epsilon - q C_0)}{H}), \quad
L_Q=\sqrt{1-4 Q^2 q^2 + 4 \l_{\alpha} (\l_{\alpha}+1)} \\ \\
\mu = \Tilde{M} - i \Tilde{\epsilon}.
\end{matrix}
\end{equation}

Here $\Tilde{M} = \sqrt{9H^2 + 4M^2}$ and $\Tilde{\epsilon} = (HQq - \epsilon + q C_0)$. A reasonable assumption is taking $C_0 = \frac{\epsilon}{q} - H Q$ since any $C_0$ shift is ineffective on wave solution observables besides energy eigenvalue shifts. This makes $\Tilde{\epsilon}$ to vanish and $\mu$ to be purely real and we have

\begin{equation} \label{KGQ_general_solutions}
\resizebox{\textwidth}{!}{%
$\begin{matrix}
\phi_1 (r) = f_{1}(r) \mathcal{H_G} \bigg(-1, \frac{(E_Q-1)  (1 + L_Q) -  4 Q^2 q^2}{-2} , \frac{(E_Q + L_Q + 1)}{2} + \frac{\Tilde{M}}{2 H},  \frac{(E_Q + L_Q + 1)}{2} - \frac{\Tilde{M}}{2 H}, 1 + L_Q, E_Q, -Hr \bigg) \\
\phi_2 (r) = f_{2}(r) \mathcal{H_G} \bigg(-1, \frac{(E_Q-1)  (1 - L_Q) -  4 Q^2 q^2}{-2} , \frac{(E_Q - L_Q + 1)}{2} + \frac{\Tilde{M}}{2 H},  \frac{(E_Q - L_Q + 1)}{2} - \frac{\Tilde{M}}{2 H}, 1 - L_Q, E_Q, -Hr \bigg) \\

\end{matrix}$
}
\end{equation}

$\frac{L_Q}{2}$ is usually denoted as critical parameter such that the possibility of particle creation depends on whether it is real or not. Note that there is a mirror symmetry of $L_Q$ between the solutions.   And $L_Q$ is purely imaginary when $|2 Q q| > |1 + 2 l_\alpha|$ which is the most expected case (See also Subsection~\ref{subsec:neutral} for the contrary case). So parameter shifts of $ L_Q \leftrightarrow - L_Q$ and $E_Q \leftrightarrow \ (2- E_Q)$ enable a transformation between complex conjugates of $\mathcal{H_G}$ functions. $ L_Q \leftrightarrow - L_Q$ is automatically satisfied due to the mirror symmetry. Now, we can implement one of the parameter shifting transformations of Heun functions to the first solution:

\begin{equation}
\resizebox{\textwidth}{!}{%
$\begin{matrix}
\mathcal{H_G}(a, q_H, \varrho, \beta, \gamma, \delta, z) = (1 - z)^{1 - \delta} \mathcal{H_G}\left(a, q_H - (\delta - 1)\gamma a,  \varrho - \delta + 1,\beta - \delta + 1, \gamma, 2 - \delta, z\right)
\end{matrix}$
}
\end{equation}
\noindent to reveal

\begin{equation}
\phi_1 (r) \, (-1)^{1-E_Q}  = r^{\frac{L_Q + L_Q^*}{2}} \, \phi_2^{*}(r). \\
\end{equation}
Keeping in mind this relation let us focus on two other properties of $\mathcal{H_G}$
\begin{equation}
\mathcal{H_G}(-1, -q_H, \varrho, \beta, \gamma, \delta, -z) = \mathcal{H_G}(1, q_H, \varrho, \beta, \gamma, \delta, z)
\end{equation}

and

\begin{equation}
\begin{matrix}
\mathcal{H_G}\! (1 ,q_H ,\varrho ,\beta ,\gamma ,\delta ,z ) = \frac{_{2}F_{1} (a,b;\gamma ;z )}{(z -1)^{(\varrho - a)} {\mathrm e}^{i \pi (b - \beta)}},
\end{matrix}
\end{equation}

where
\begin{equation}
\begin{matrix}
a=\frac{\sqrt{\gamma^{2}+2 (-\varrho -\beta ) \gamma +(\varrho -\beta )^{2}+4 q_H}}{2}+\frac{\gamma}{2}+\frac{\varrho}{2}-\frac{\beta}{2}, \\
b= \frac{\sqrt{\gamma^{2}+2 (-\varrho -\beta ) \gamma +(\varrho -\beta )^{2}+4 q_H}}{2}+\frac{\gamma}{2}-\frac{\varrho}{2}+\frac{\beta}{2}
\end{matrix}
\end{equation}

Thus, wave solutions can be reduced to generalized hypergeometric functions such as

\begin{equation}
\begin{matrix}
\phi_1 (r) = f_{1}(r) (Hr -1)^{(a - \varrho)} e^{i \pi (\varrho - a)} \, _{2}F_{1} (a,b;1 + L_Q ; H r )   \\
\end{matrix}
\end{equation}

Generalized hypergeometric functions have properties given below that will also be used later in this paper:

\begin{equation} \label{relations}
\resizebox{\textwidth}{!}{%
$\begin{matrix}
_{2}F_1(a,b;c;z) = \, _{2}F_1(b,a;c;z)  \\ \\
_{2}F_1(a,b;c;z) = (1-z)^{c-a-b} \, _{2}F_1(c-a, c-b; c; z)\\ \\
_{2}F_1(a, b; c; z) = \frac{\Gamma(c) \Gamma(c-a-b) _{2}F_1(a, b; a+b+1-c; 1-z)}{\Gamma(c-a) \Gamma(c-b)} \,  +
\frac{ (1-z)^{c-a-b} \Gamma(c) \Gamma(a+b-c) _{2}F_1(c-a, c-b; 1+c-a-b; 1-z)}{\Gamma(a) \Gamma(b)}  \,
\end{matrix}$
}
\end{equation}

\noindent where the first relation is the symmetry relation, the second relation is called Euler Transformation and the third one is one of the 24 connection formulae of hypergeometric functions. Thus, one can construct a third solution of the wave equation, $\phi_3(r)$, choosing the parameter set of  $c=1+a+b-\gamma$, $z= 1 - Hr$ and inserting on the third line of  \ref{relations} and one gets:

\begin{equation} \label{Phi3(r)}
\begin{matrix}
\phi_3(r) = N \frac{\Gamma(c) \Gamma(c-a-b)}{\Gamma(c-a) \Gamma(c-b)} \, \phi_1(r) + N
\frac{  \Gamma(c) \Gamma(a+b-c) H^{-L_Q} (-1)^{(E_Q - 1)} }{\Gamma(a) \Gamma(b)} \phi^*_1(r)  \,
\end{matrix}
\end{equation}

A physically meaningful explanation can be provided using Bogoliubov transformations in such cases \cite{Bogoliubov1, Bogoliubov2}, wherein creation and annihilation operators describe changes in particle states across different reference frames. The observation of vacuum may be relative such that the definition differs from different observers. The transformation introduces new states that represent a new particle content, leading to effects like particle creation from vacuum fluctuations. If we place an observer far from the event horizon and denote the observed wave as  $\phi^{+}_{out}$, we can express it as follows:
\begin{equation} \label{Bogoliubov}
\begin{matrix}
\phi^{+}_{out} = A \phi^{+}_{in}  + B \phi^{-}_{in} \\
\phi^{-}_{in}=[\phi^{+}_{in}]^*
\end{matrix}
\end{equation}

\noindent where $\phi_{in}$ states denote the created particle waves around the horizon. A comparison between Eq. (21) and (22) shows that $\phi_1, \phi_2 \sim \Bar{\phi}_1, \phi_3$ correspond to $\phi_{in}^+, \phi_{in}^-, \phi_{out}^+$, respectively. $N$ in  (\ref{Phi3(r)}) is for normalization constraint such that $|A|^2 - |B|^2 = 1$  is satisfied. Thus, we have particle creation probability as a distribution of particle energy and charge

\begin{equation} \label{subsec1_B/A}
\begin{matrix}
P=|\frac{B}{A}|^2 = -1^{2(E_Q-1)} = e^{- 2 \pi (Qq + \frac{\epsilon - q C_0}{H}) }
\end{matrix}
\end{equation}

\begin{figure}[H]
    \centering
    \includegraphics[width=0.8\textwidth]{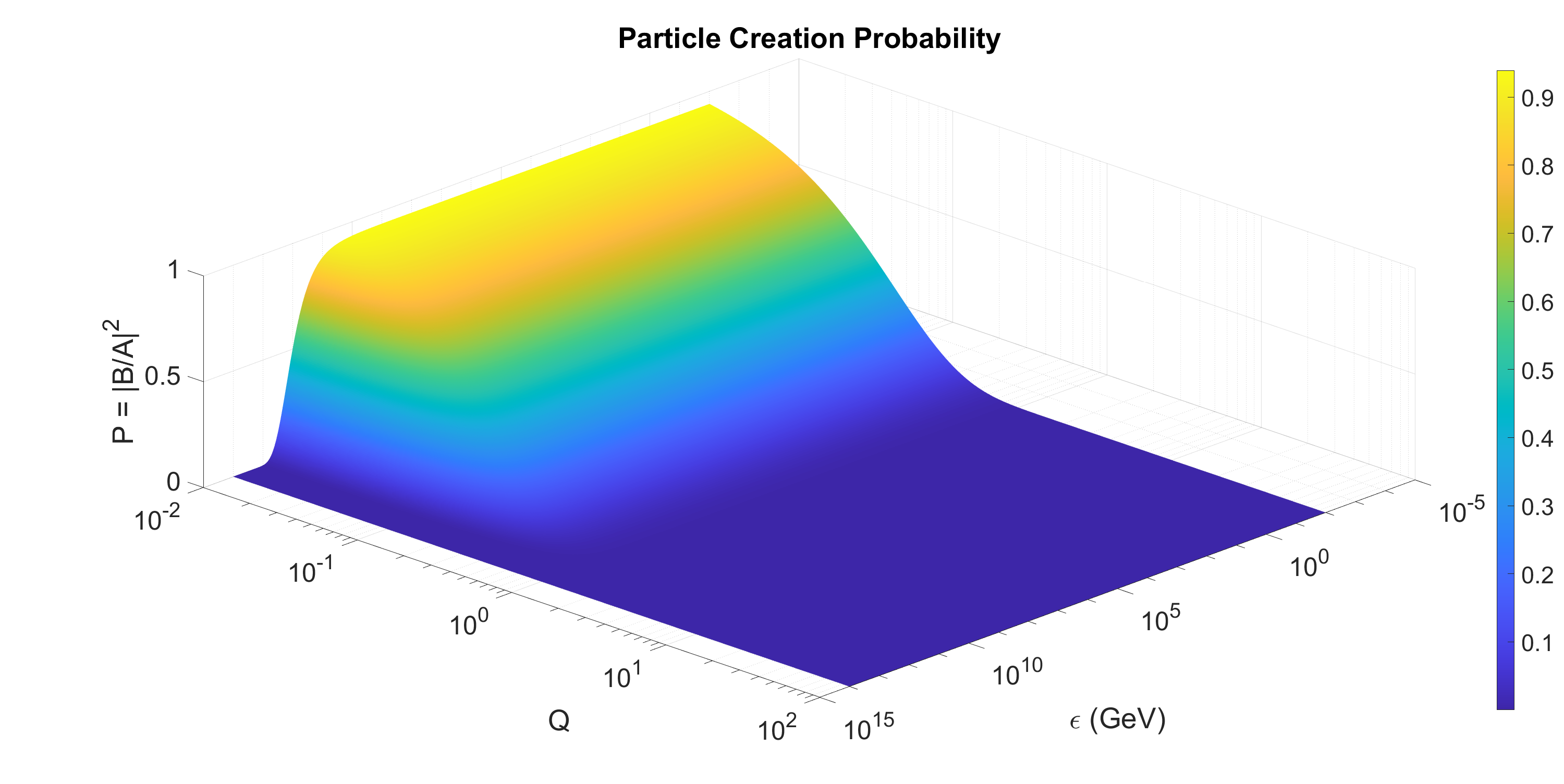} % Include the EPS file
        \caption{Particle creation probability $|\frac{B}{A}|^2$ regarding the exact solution of point-like source ($C_0=0$,\,$H=10^{13} GeV$). $Q$ is given in natural units. Probability arises while the energy eigenvalue of created particles decreases which is in conformity with astrophysical observations.}
    \label{fig:1}
\end{figure}

\begin{figure}[H]
    \centering
    \includegraphics[width=0.8\textwidth]{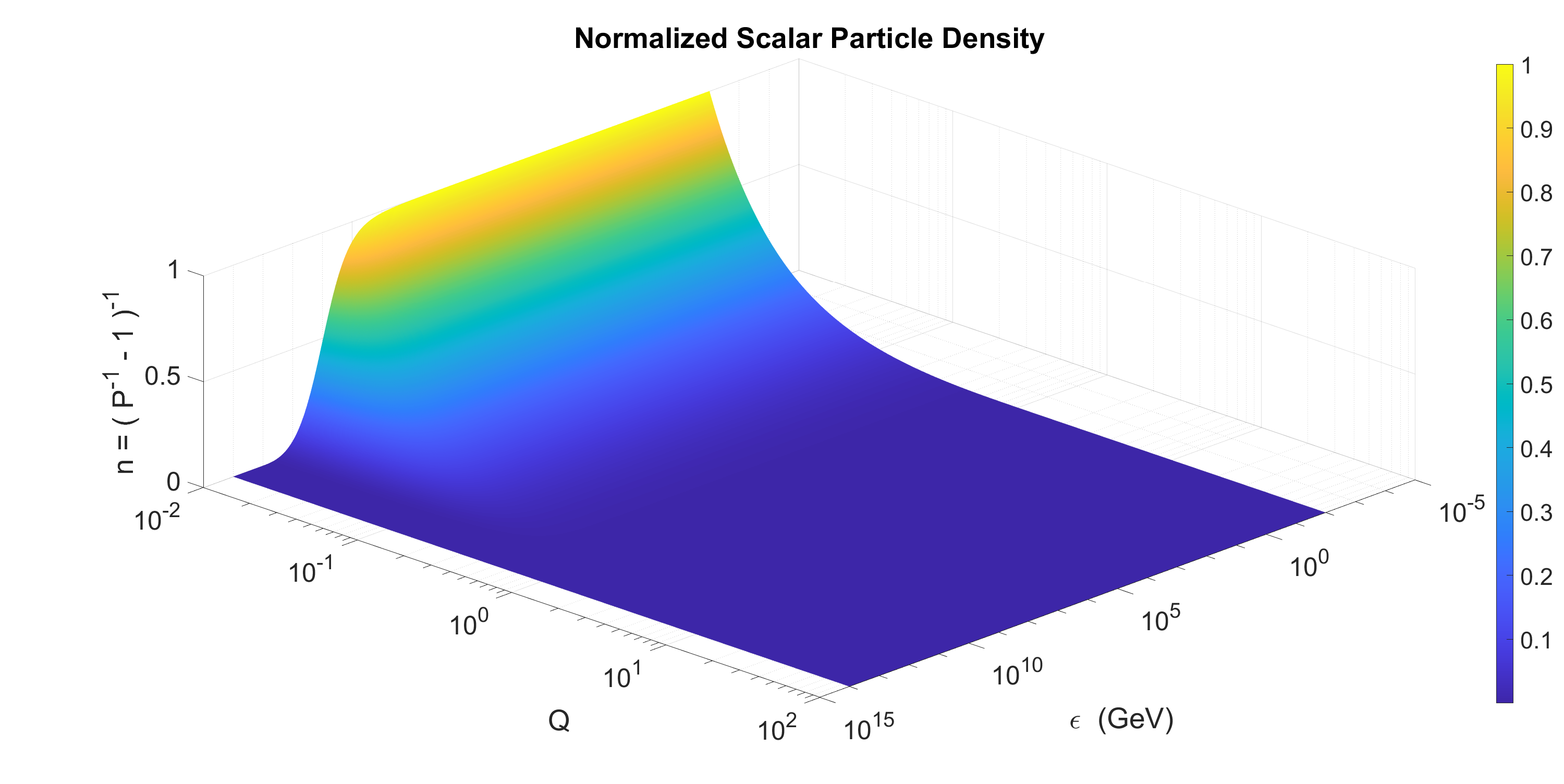} % Include the EPS file
    \caption{Scalar particle density, $n = \frac{1}{P^{-1} - 1}$ regarding the exact solution of point-like source ($C_0=0$,\,$H=10^{13} GeV$). $Q$ is given in natural units. Spectrum and density of emitted particles could provide insights into the properties of cosmic string charge.}
    \label{fig:2}
\end{figure}

Although the results of the limits of $Q \to 0$ and $Q \to \infty $ can easily be seen from  (\ref{subsec1_B/A}), we continue with detailed analyses of these limits to reveal an interesting relationship between them.

\section{Other Cases}

\subsection{Cosmic String with Dense Point-like Source} \label{Dense}

Now let us consider that electrically dense point source ($Q \gg 1$) is trapped by a cosmic string in a space-time given in the same metric, $g$, as given in the previous Section. This choice modifies the radial part of the Klein-Gordon equation as shown below, while leaving the angular part unchanged,

\begin{equation} \label{KGQ}
\resizebox{\textwidth}{!}{%
$\begin{aligned}
(1-H^2r^2) \phi(r)^{''} + \frac{2}{r} (1-2H^2r^2) \phi(r)^{'} +
\bigg(\frac{\epsilon^2 + (Q q / r)^2}{1-H^2r^2} + M^2 - \frac{l_{\alpha} (l_{\alpha}+1)}{r^2}  \bigg) \phi(r) = 0.
\end{aligned}$
}
\end{equation}

Here we assumed $C_0 = 0$ and $Q q \gg 2 \epsilon r$ throughout the static de Sitter space ($0<r<1/H$). The solution for this equation is given as

\begin{equation} \label{Phi}
\begin{matrix}
\phi_1 (r) = r^{\frac{-1}{2}+L'_Q} (H^{2} r^{2}-1)^{-E'_Q}  \hspace{2pt} _{2}F_{1}\! (\zeta, \xi; 1+L'_Q ;H^{2} r^{2}),
\\
or
\\
\phi_2 (r) = r^{\frac{-1}{2}-L'_Q} (H^{2} r^{2}-1)^{-E'_Q}  \hspace{2pt} _{2}F_{1}\! (\kappa, \sigma; 1-L'_Q ;H^{2} r^{2})
\end{matrix}
\end{equation}

\noindent or any linear combinations of them where we redefine the parameters such as

\begin{equation} \label{terimler}
\begin{matrix}
\zeta = \frac{1+L'_Q}{2} - E'_Q - \Tilde{M'}  \\
\xi = \frac{1+L'_Q}{2} - E'_Q + \Tilde{M'} \\
\kappa = \frac{1-L'_Q}{2} - E'_Q + \Tilde{M'} \\
\sigma = \frac{1-L'_Q}{2} - E'_Q - \Tilde{M'} \\
\\
L'_Q= \frac{L_Q}{2}=\frac{\sqrt{1-4 Q^2 q^2 + 4 \l_{\alpha} (\l_{\alpha}+1)}}{2}  \quad E'_Q=\frac{\sqrt{-Q^2 q^2 H^2 - \epsilon^2}}{2H} \quad \Tilde{M'}=\frac{\sqrt{9H^2 + 4M^2}}{4H}
\end{matrix}
\end{equation}

Let us analyze asymptotical behaviour of these waves to understand their characteristics. Hypergeometric functions of the solutions always tend to stay constant at limits since

\begin{equation} \label{limitler}
\begin{matrix}
\lim_{z \to 1} \quad {}_2F_1(\frac{1}{2} + m - k, -n; 2m + 1; z ) \to \frac{\Gamma(2m + 1) \Gamma\left(m + \frac{1}{2} + k + n\right)}{\Gamma\left(m + \frac{1}{2} + k\right) \Gamma(2m + 1 + n)} \\
\lim_{z \to 0} \quad {}_2F_1(m,n;k;z) \to 1.
\end{matrix}
\end{equation}

Thus we have

\begin{equation} \label{limitler2}
\begin{matrix}
\lim_{r \to 0} \quad \phi_1 (r) \to r^{\frac{-1}{2}+L'_Q}  \\
\lim_{r \to 1/H} \quad \phi_1 (r) \to (H^{2} r^{2}-1)^{-E'_Q} \\
\lim_{r \to 0} \quad \phi_2 (r) \to r^{\frac{-1}{2}-L'_Q}   \\
\lim_{r \to 1/H} \quad \phi_2 (r) \to (H^{2} r^{2}-1)^{-E'_Q}
\end{matrix}
\end{equation}

Our assumption, as presented in the beginning of this subsection, enables us to consider $L'_Q$ and $E'_Q$ to be purely imaginary. This directly brings an oscillatory behaviour by phases of $\pm \alpha \, ln(r)$. Thus, $\phi_1 (r)$ acts like an outgoing wave while $\phi_2 (r)$ as an incoming one.
 One can easily see that a suitable choice of the parameters for the connection formulae given in the third line of Eq. \ref{relations} may give:

\begin{eqnarray}\label{connection1}
% \nonumber to remove numbering (before each equation)
  _{2}F_1(\zeta, \xi; 1-2E'_Q; 1-H^2 r^2) =  \frac{\Gamma(1-2E'_Q) \Gamma(-L'_Q) \; _{2}F_1(\zeta, \xi; 1+L'_Q; H^2 r^2)}{\Gamma(\sigma) \Gamma(\kappa)} + \\
    \frac{ (H^2 r^2)^{-L'_Q} \Gamma(1-2E'_Q) \Gamma(L'_Q) \; _{2}F_1(\kappa, \sigma; 1-L'_Q; H^2 r^2)}{\Gamma(\zeta) \Gamma(\xi)}  \nonumber
\end{eqnarray}
Multiplying both sides by $(H^2 r^2 - 1)^{-E'_Q} r^{\frac{-1}{2} + L'_Q}$ gives

\begin{equation} \label{connection2}
\begin{aligned}
\phi_3(r) =  \frac{\Gamma(1-2E'_Q) \Gamma(-L'_Q)} {\Gamma(\sigma) \Gamma(\kappa)} \, \phi_1(r)  +
\frac{\Gamma(1-2E'_Q) \Gamma(L'_Q)}{\Gamma(\zeta) \Gamma(\xi)} H^{-2 L'_Q} \, \phi_2(r)  \,
\end{aligned}
\end{equation}

where

\begin{equation} \label{Phi3}
\begin{aligned}
\phi_3(r) = (H^2 r^2 - 1)^{-E'_Q} \, r^{\frac{-1}{2} + L'_Q} \, _{2}F_1(\zeta, \xi; 1-2E'_Q; 1-H^2 r^2)
\end{aligned}
\end{equation}

(Please see Appendix-B for a detailed analysis on $\phi_3(r)$). On the other hand, applying Euler transformation to $\phi_2(r)$ shows that

\begin{equation} \label{Phi2}
\begin{aligned}
\phi_2(r) = (-1^{-2 E'_Q}) \phi_1^*(r)
\end{aligned}
\end{equation}

 Inserting Eq. \ref{Phi2} into Eq. \ref{connection2} we have

\begin{equation} \label{B/A}
\lim_{Q \to \infty} P=|\frac{B}{A} |^2 = e^{- 4 \pi |E'_Q|}
\end{equation}

The result becomes $e^{- 2 \pi Q q}$ for large $Q$ which is consistent with the results obtained in the previous Section.

\subsection{Electrically Neutral Cosmic String} \label{subsec:neutral}

Another possible physical situation is the cosmic string to be electrically neutral. This corresponds to easily taking $Q \to 0$ on our findings. However, this limit deserves special attention since the KG solutions, their behaviours and physical meanings are directly changed by it. The main structure of the wave equation and corresponding wave solutions remain the same as in Section \ref{Dense}, but this time we have

\begin{equation} \label{yeniTerimler}
\begin{matrix}
L''_Q=\frac{1}{2} + l_\alpha  \quad E''_Q=\frac{i \epsilon}{2H} \quad \Tilde{M''}=\Tilde{M'}.
\end{matrix}
\end{equation}

\noindent
In other words $E''_Q$ remains purely imaginary while this time $L'_Q$ becomes purely real. $\phi_1(r)$ and $\phi_2(r)$ are no longer complex conjugate couples in this case. So, interestingly, the Bogoliubov transformation should be set by different combinations of incoming and outgoing waves. First let us implement connection formula to $\phi_1(r)$  to get

\begin{equation} \label{Phi1Connection}
\begin{aligned}
\phi_1(r) =  \frac{\Gamma(1+L''_Q) \Gamma(2 E''_Q) } {\Gamma(\zeta^*) \Gamma(\xi^*)} \, \phi_3(r)  +
\frac{\Gamma(1+L''_Q) \Gamma(-2 E''_Q)}{\Gamma(\zeta) \Gamma(\xi)} (-1^{2E''_Q}) \phi_3^*(r) \,
\end{aligned}
\end{equation}

In the neutral case $\phi_1(r)$ is $\phi^+_{out}$ while $\phi_3(r)$ and its complex conjugate are the input states. Therefore, now we have

\begin{equation} \label{B/A yeni}
\lim_{Q \to 0} P = |\frac{B}{A}|^2 = e^{- 4 \pi |E''_Q|} = e^{- 2 \pi \epsilon / H}.
\end{equation}

This result shows that there is a balance between created and annihilated particles for $H  \gg  \epsilon$. It implies that for each particle created, there is an equal probability of the corresponding particle being annihilated. In other words, the process near the cosmic string exhibits symmetric particle creation and annihilation. From the perspective of quantum mechanics, this symmetry ensures conservation of probability, which aligns with the principle of unitarity \cite{hawking}. The unitarity condition is essential to maintaining a consistent quantum theory, where no information is lost during the particle creation and annihilation processes. One can read the related study about the non-uniqueness of the vacuum state and how different observers may disagree on whether particles are present, which is key to understanding particle creation in \cite{fulling}. Moreover, this result highlights the intriguing property of vacuum state non-uniqueness in curved spacetime. Different observers, particularly those at varying distances or in different frames of reference, may not agree on whether particles are present \cite{fulling}. This ambiguity lies at the heart of understanding particle creation in curved spacetimes and underscores the richness of quantum field theory in the presence of gravitational effects.

Here, the number of particles created  \emph{N}  is related to the modulus of the Bogoliubov coefficient $B$  which corresponds to the creation of particles. Then, since $|A|^2 - |B|^2 = 1$, it follows that:
\begin{equation}\label{4}
  \langle N \rangle= \int |B(t)|^{2} \Gamma dt = \frac{\Gamma \Delta t}{P^{-1} - 1} = (e^{ 2 \pi \frac{\epsilon}{H}} - 1)^{-1} \Gamma \Delta t
\end{equation}
\noindent assuming $B$ and product rate $\Gamma \propto \frac{2 \pi G_{\mu}}{\epsilon}$ are time invariant. For example; if we calculate  \emph{N} for a time interval of one year (approximately $4.8 \times 10^{31} GeV^{-1}$) during the early universe in which energy scale can be taken in GUT scale as $\epsilon=4.0 \times 10^{13} GeV$ and Hubble constant as $H=10^{13} GeV$ then expected value of number of created scalar particles $\langle N \rangle$ is found to be 92 for $G_{\mu} = 10^{-6}$.

We also give number of created scalar particle distribution with respect to energy in Figure 3.

\begin{figure}[H]
    \centering
    \includegraphics[width=0.8\textwidth]{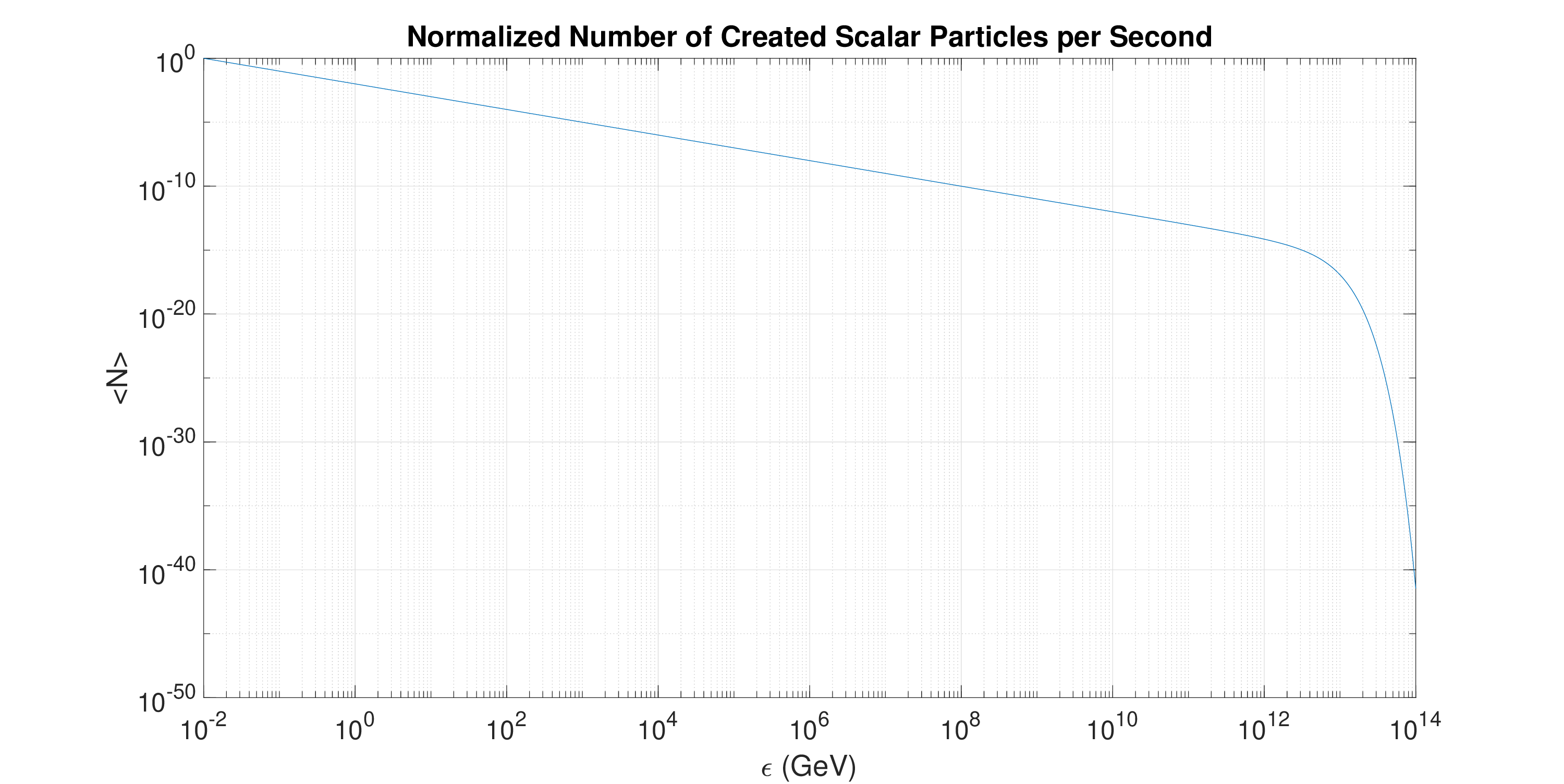} % Include the EPS file
    \caption{Expected value comparison of number of created scalar particles, $\langle N \rangle$,  within  a second ($1.52 \times 10^{24} GeV^{-1}$) assuming a neutral cosmic string. The plot is given after normalization and the calculations are done for $\pm 1 \%$ energy interval around the target energy.}
    \label{fig:3}
\end{figure}

\subsection{Solution for the Linear Potential}

Another intriguing problem is a cosmic string with linear potential in de Sitter space $V(r) = \frac{V_0}{d} r - C_0$ where $V_0 = \frac{Q'}{r_0}$. This scenario may occur when an external field within a local spatial potential difference, $\frac{Q'}{r_0}$, exists and a cosmic string is located in that part of the space. Since $r_0$ and $d$ are constants in terms of length in natural units we may take $Q_l(r) = \frac{Q'}{r_0 d}$ and insert necessary unit conversions later. This time, the wave equation becomes

\begin{equation} \label{KGQ_linear}
\resizebox{\textwidth}{!}{%
$\begin{aligned}
(1-H^2r^2) \phi(r)^{''} + \frac{2}{r} (1-2H^2r^2) \phi(r)^{'} +
\bigg(\frac{(\epsilon - q Q_l r - q C_0)^2 }{1-H^2r^2} + M^2 - \frac{l_{\alpha} (l_{\alpha}+1)}{r^2}  \bigg) \phi(r) = 0.
\end{aligned}$
}
\end{equation}

The solution almost keeps its exact form as in the point-like source

\begin{equation}
\resizebox{0.95\textwidth}{!}{%
$
\begin{matrix}
\phi_1 (r) = f_{1}(r) \mathcal{H_G} \bigg(-1, \frac{(E_Q-1)  (1 + L_Q)}{-2} + T_1, \frac{H (E_Q + L_Q + 1) + \mu}{2 H}, \frac{H (E_Q + L_Q + 1) \mu - (\Tilde{M}^2 + \Tilde{\epsilon}^2)}{2 \mu H}, 1 + L_Q, E_Q, -Hr \bigg) \\
\phi_2 (r) = f_{2}(r) \mathcal{H_G} \bigg(-1, \frac{(E_Q-1)  (1 - L_Q)}{-2} + T_2, \frac{H (E_Q - L_Q + 1) + \mu}{2 H}, \frac{H (E_Q - L_Q + 1) \mu  - (\Tilde{M}^2 + \Tilde{\epsilon}^2)}{2 \mu H}, 1-L_Q, E_Q, -Hr \bigg)
\end{matrix}$
}
\end{equation}

\noindent with redefined parameters:
\begin{equation}
\begin{matrix}
T_1 = i \frac{  (1 + L_Q) \Tilde{\epsilon} \, \mu^{*}}{ 2 \mu H}, \quad T_2 = i \frac{  (1 - L_Q) \Tilde{\epsilon} \, \mu^{*}}{ 2 \mu H} \\
E_Q = 1 + i( \frac{Q_l q}{H^2} + \frac{\epsilon - q C_0}{H}), \quad
L_Q=\sqrt{1 + 4 \l_{\alpha} (\l_{\alpha}+1)} \\ \\
\end{matrix}
\end{equation}

Here also $\Tilde{M} = \sqrt{9H^2 + 4M^2 - \frac{4 Q_l^2 q^2}{H^2}}$ and $\Tilde{\epsilon} = (\frac{Q_l q}{H} - \epsilon + q C_0)$ are redefined while rest of the parameter definitions are kept the same as in Section \ref{Exact_pointlike}. $\Tilde{M}$ can be assumed to be real when $\frac{Q_l q}{H^2}$ compared to $H^2$ and choice of $C_0 = \frac{\epsilon H - Q_l q}{q H}$ vanishes $\Tilde{\epsilon}$. New solutions become as below
\begin{equation} \label{KGQ_general_solutions}
\resizebox{0.95\textwidth}{!}{%
$\begin{matrix}
\phi_1 (r) = f_{1}(r) \mathcal{H_G} \bigg(-1, \frac{(E_Q-1)  (1 + L_Q)}{-2} , \frac{(E_Q + L_Q + 1)}{2} + \frac{\Tilde{M}}{2 H},  \frac{(E_Q + L_Q + 1)}{2} - \frac{\Tilde{M}}{2 H}, 1 + L_Q, E_Q, -Hr \bigg) \\
\phi_2 (r) = f_{2}(r) \mathcal{H_G} \bigg(-1, \frac{(E_Q-1)  (1 - L_Q)}{-2} , \frac{(E_Q - L_Q + 1)}{2} + \frac{\Tilde{M}}{2 H},  \frac{(E_Q - L_Q + 1)}{2} - \frac{\Tilde{M}}{2 H}, 1 - L_Q, E_Q, -Hr \bigg) \\

\end{matrix}$
}
\end{equation}

Applying the same procedure to $\phi(r)$ functions we obtain

\begin{equation} \label{B/A yeni}
P = |\frac{B}{A}|^2 = e^{- 2 \pi ( \frac{Q_l q}{H^2} + \frac{\epsilon - q C_0}{H})}
\end{equation}

In (\ref{B/A yeni}), it shows that linear potential contributes directly to the particle creation rate. A stronger potential (larger $Q_l$) or a stronger charge coupling $q$ increases $P$,  leading to more significant particle production. If $\epsilon > qC_0$, this term increases the exponent, reducing $P$  , which corresponds to fewer particles being created due to higher effective energy levels. If $\epsilon \approx q C_0$, the potential shift counterbalances the energy, simplifying the dependence.

\begin{figure}[ht]
    \centering
    \includegraphics[width=0.8\textwidth]{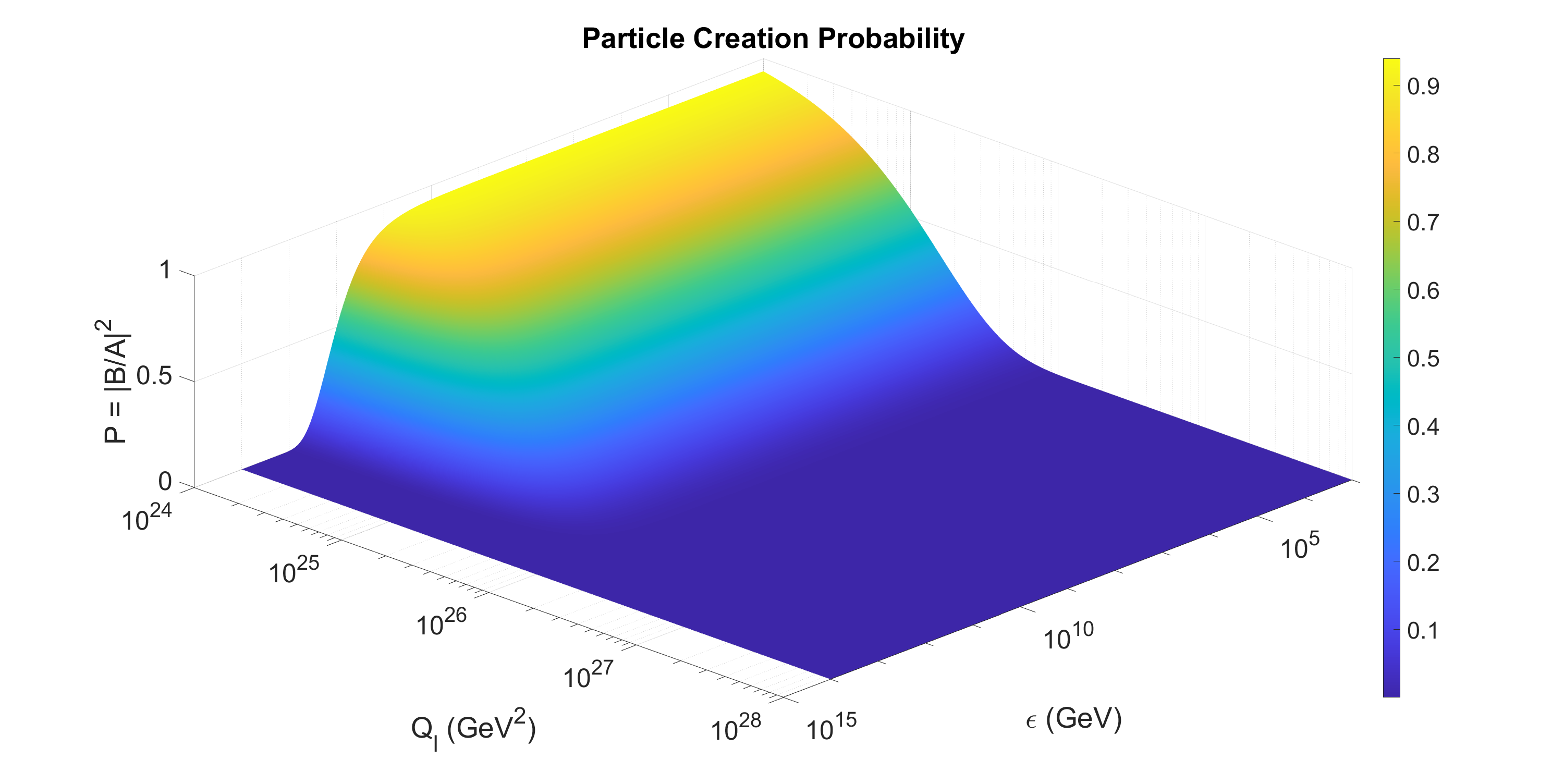} % Include the EPS file
    \caption{Particle creation probability $|\frac{B}{A}|^2$ regarding the exact solution of linear potential.}
    \label{fig:4}
\end{figure}

\begin{figure}[H]
    \centering
    \includegraphics[width=0.8\textwidth]{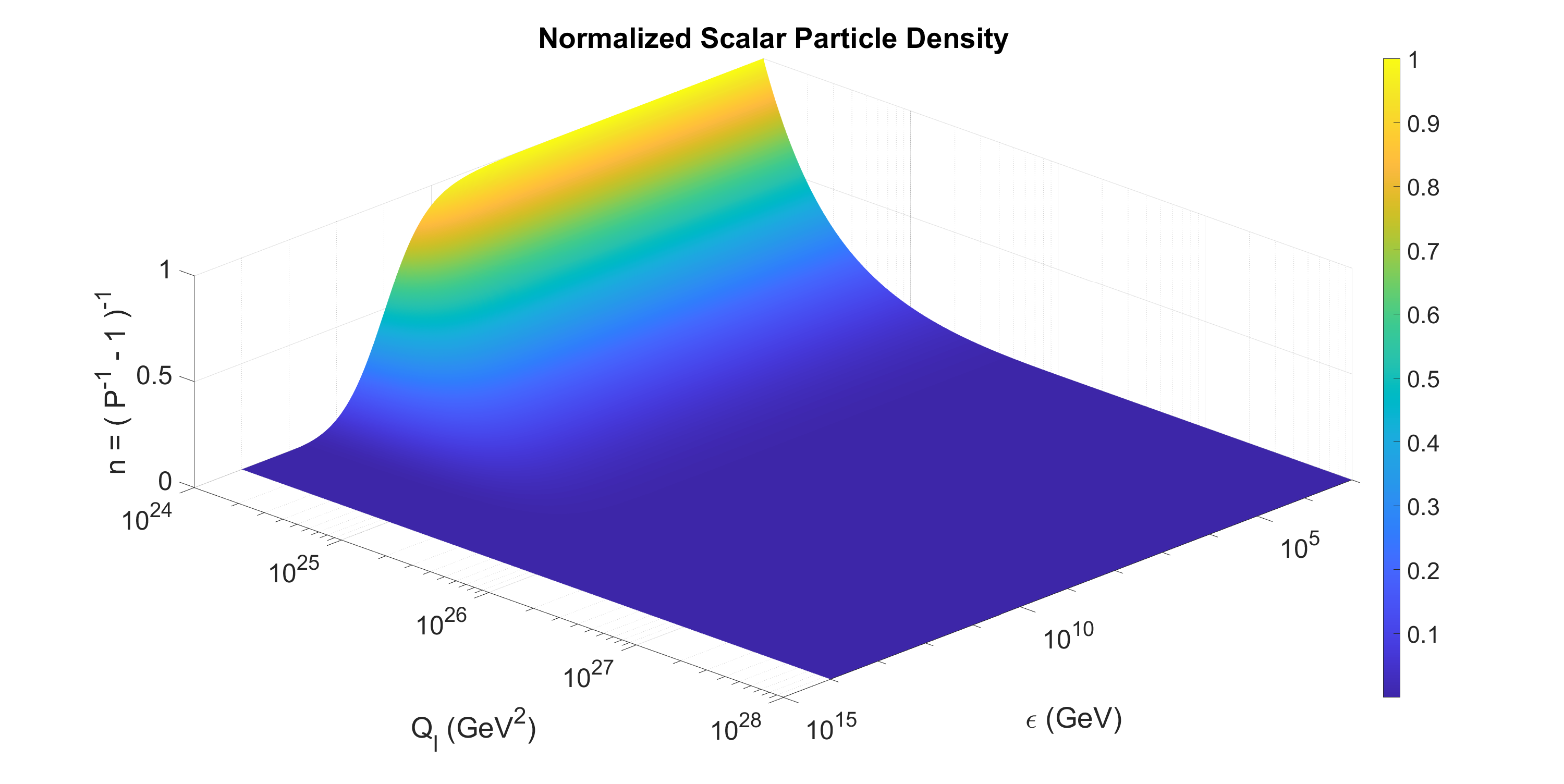} % Include the EPS file
    \caption{Scalar particle density, $n = \frac{1}{P^{-1} - 1}$ regarding the exact solution of linear potential.}
    \label{fig:5}
\end{figure}

\newpage
\section{Conclusions}
In this paper, we have studied the Klein-Gordon (KG) equation for a cosmic string in de Sitter spacetime, focusing on scalar fields as a baseline for understanding quantum particle creation in curved spacetime. By solving the KG equation, we derived the radial and angular wave equations, identifying key features of the effective potential, which incorporates contributions from the angular deficit of the cosmic string and the curvature of de Sitter spacetime. Our analysis reveals that particle creation probability, calculated from the effective potential and Bogoliubov coefficients, exhibits an exponential dependence on particle energy.

This reflects the tunneling probability across the potential barrier created by the angular deficit $\alpha$ and Hubble parameter $H$.  Larger $H$  enhances the particle creation probability for low-energy modes, while a smaller $\alpha$ increases the angular momentum term, suppressing particle creation for higher angular quantum numbers. A heavier scalar field introduces a higher energy threshold, reducing particle creation probability for given modes.

It is shown for the first time that distinct charge limits of a cosmic string in an expanding space lead to a huge differentiation on supercharge interpretations by the help of Bogoliubov transformations. Equation (36) shows that there is possibility of particle creation under supercritical source charge $Qq \rightarrow 0$ due to expansion of space. The corresponding neutral pseudoscalar solutions are shown to be related to complex solutions of over supercritical source $Qq > \frac{1}{2}+l_{\alpha}$. The neutral and static space limits $(Q \rightarrow 0 \, \And \, H \rightarrow 0)$ vanishes the particle creation probability which is in conformity with the literature\cite{Belbaki}.

Figures 1 and 2 illustrate the particle creation probability and scalar particle density for a dense point-like system: in Figure 1, for small $Q (Q<10^{1})$, the particle creation probability is high  $ \sim 1$, indicating efficient tunneling and particle creation in this regime. For large $Q(Q>10^{3})$, the probability drops significantly, approaching values near $0.1$ or lower. This suppression is consistent with the increased potential barrier caused by larger $Q$, which reduces tunneling probabilities. When we look at the energy dependence $\epsilon (\epsilon < 10^{0}) GeV$, the creation probability is relatively high across all $Q$, showing a strong preference for low-energy particle production. For large $\epsilon(\epsilon > 10^{5}) GeV$, the probability drops off sharply, consistent with the exponential suppression term $e^{-2\pi \epsilon /T}$  in thermal-like particle creation processes. At intermediate values of  $Q$ and $\epsilon$, the creation probability transitions smoothly from high to low values, forming a "ridge" in the $Q-\epsilon$ space where particle creation is most likely. In Figure 2, the scalar particle density $n$ is high, with values close to $1$, indicating efficient particle production in this regime. The density decreases significantly for the larger values of $Q$, reflecting the reduced particle creation probability due to the higher potential barrier associated with larger charges. The scalar particle density remains high across all $Q$, showing a preference for low-energy particle production. On the other hand, The density drops sharply, as higher-energy particles are suppressed by the exponential term in the particle creation probability. The Figure 3 presents the expected number of scalar particles $N$ created per second as a function of the energy,  assuming a neutral cosmic string. When $\epsilon < 10^{3} GeV$, and $G_\mu = 10^{-6}$  the number of created particles  is extremely high, reaching values above $10^{30} s^{-1}$. This shows that low-energy modes dominate the particle production process, consistent with thermal-like quantum field behavior. Beyond $\epsilon > 10^{12} GeV$, the number of created particles $N$ decreases sharply, approaching $10^{-6} s^{-1}$. In the transition region $10^{3} < \epsilon < 10^{12} GeV$,  the particle creation rate transitions smoothly from high to low values. Figure 4 illustrates the particle creation probability as a function of the charge parameter and energy  based on the exact solution of a linear potential. For the values of $Q_l \sim 10^{24} GeV^{2}$, the particle creation probability is near its maximum ($\sim 1$) , indicating efficient creation of scalar particles in this regime. For $Q_l \sim 10^{28} GeV^{2}$, the creation probability decreases significantly, dropping below $0.1$, reflecting the suppression caused by the linear potential's barrier as $Q$ increases. For low energies $\epsilon < 10^{5} GeV$, the particle creation probability is high across a wide range of $Q$, showing a preference for low-energy modes. For $\epsilon > 10 ^{10} GeV$, the creation probability declines sharply, consistent with the exponential suppression in thermal-like systems and the increased barrier height in the linear potential at higher energies. Figure 5  the normalized scalar particle density  $n$ with a linear potential. Low region $Q_l (Q_l \sim 10^{24}) GeV^{2}$, the scalar particle density $n$ is high, reaching values near $1$, indicating efficient scalar field population in this regime. For $Q_l \sim 10^{28} GeV^{2}$ The density sharply decreases, dropping below $0.2$ reflecting the suppression of particle creation due to the stronger linear potential barrier as $Q$ increases. In the low energy region $\epsilon < 10^{5} GeV$, the scalar density remains relatively high, showing a preference for low-energy particle production across all $Q$. In the high energy region $\epsilon > 10^{10} GeV$, the density drops steeply, with $n$ falling below $0.1$, consistent with the exponential suppression of high-energy particle creation probabilities.

The point-like system represents cases where particle creation is strongly localized near the source $r\rightarrow 0$, enabling efficient particle creation over a broader range of $Q$ and $\epsilon$.  In contrast, the linear potential introduces extended confinement, limiting particle creation to low
$Q$ and $\epsilon$.  While the point charge potential allows for higher scalar densities and particle production efficiency, the linear potential reflects a more restrictive environment, suppressing scalar particle densities and creation probabilities at higher charges and energies.

The limit $H\rightarrow0$ is the static space limit where only cosmic string effects are expected on the metric and therefore on the results (The related parameter is $\alpha$ in this study). This case has already been studied and given in Ref. \cite{Belbaki}. Since angular wave equation is independent of $H$, one expects it to be the same as given in Ref. \cite{Belbaki} which is also mentioned in the paper before Eq. (7). However, radial equation depends on both $\alpha$ (explicitly $l_\alpha$) and $H$ and therefore totally differs from prior findings and the difference of these studies rise here.
In our findings we also reach a result that it is possible to investigate $L_Q=2\sqrt{(l_{\alpha}+1/2)^2-(qQ)^2}$ in two different cases where it is imaginary or real. Here $L_Q/2$ directly corresponds to critical parameter explained in Introduction of \cite{Duru} and in Chapter 7 of \cite{QED} which defines supercritical source as $Qq>l_\alpha+1/2$. This condition gives a mixture of two complex conjugate states as we arrived in the Section 3.1. So our findings match with the literature. However we also show that when $Q$ is under supercritical condition ($L_Q/2$ is purely real) there are still wavefunction solutions that can be handled under pair production mechanism and these solutions still depend on the charged string solutions via Bogoliubov transformations. The production possibility arises via the expansion term $H$ which also shows itself in Eq. (36). In static space, where $H$ vanishes, the production probability due to a neutral cosmic string also vanishes. On the contrary, if spacetime expansion exists then probability is directly related with particle energy and Hubble constant. Moreover, while previous studies have analyzed quantum effects of cosmic strings and de Sitter expansion separately, our work provides a unified framework that incorporates both. Unlike past research, which focused on vacuum polarization \cite{Mello} or pair creation in static backgrounds \cite{Dias}, our study investigates particle production in a fully dynamic de Sitter background with a cosmic string. The results reveal a new mechanism where cosmic strings can alter the vacuum structure in an expanding universe, leading to modified quantum particle spectra. Additionally, our findings quantitatively generalize Hawking radiation and Unruh-like effects by introducing a string-induced energy shift in the Bogoliubov coefficient calculations.

In specific high-energy astrophysical environments, such as those surrounding cosmic strings, the production of scalar particles could leave detectable imprints \cite{imprint1, imprint2}. These imprints may include low-energy gamma rays from pion decays or contributions to the cosmic background radiation. Investigating the nature of these scalar particles and their interactions with the cosmic environment represents an exciting avenue for future research. Such efforts have the potential to connect quantum field theory predictions with observational signatures.

\newpage
\section{Appendix-A}
The Laplace and Poisson's Equations in curved spacetime are given as

\begin{equation}
  \nabla^2 A_0 = \frac{1}{\sqrt{-g}} \partial_i (\sqrt{-g}g^{ij} \partial_j A_0)
\end{equation}

\begin{equation}
  \nabla^2 A_0 = \rho = - \frac{Q}{\sqrt{-g}}\delta (r)
\end{equation}

\noindent where the upper indices are for contravariant metric tensor that corresponds to inverse of covariant metric tensor in this case. Then considering $g$ given in Eq.(3) and keeping in mind that $A_0$ is only $r$ dependent for a static point source we can write

\begin{equation}
  \partial_r(\alpha r^2 sin(\theta) (H^2 r^2 -1) \partial_r A_0 ) = 0, r \neq 0
\end{equation}

\noindent the equation can be rewritten as (multiplying both sides by constant terms)

\begin{equation}
    r^2 (1-H^2 r^2) \partial_r A_0  = C
\end{equation}

\noindent where C is a constant. Therefore

\begin{equation}
    A_0 = \int \frac{C}{r^2 (1-H^2 r^2)} \,dr =\frac{-C}{r}  + C H atanh(Hr) + C_0
\end{equation}

The particle production occurs near event horizon where $r \ll 1⁄H$  and then $atanh⁡(Hr)$ vanishes. We finally have

\begin{equation}
    A_0 = \pm \frac{C}{r} + C_0
\end{equation}

The sign of $A_0$ depends on the charge $Q$. Now derivation method of $C$ is well-known such that Gauss’ Law is applied near the proximity of $r=0$

\begin{equation}
\oint E \,d\Sigma = \frac{Q}{\epsilon_0^{eff}}
\end{equation}

\noindent where $E= - \partial_r A_0(r)$, $\epsilon_0^{eff}$ is vacuum permittivity. There is a possibility of permittivity to be different from that of Minkowski due to both curvature of de Sitter and gravitational effects. It is still a constant within the surface are considered in the integral. Differential surface area $d \Sigma$ is given by

\begin{equation}
d \Sigma = \sqrt{g_{\theta \theta} g_{\varphi \varphi}} d \theta d \varphi = \alpha r^2 sin(\theta) d \theta d\varphi
\end{equation}

\noindent to finally find out $C \sim Q$ in natural units. This result shows that it is reasonable to consider point-source as given in Eq. (9).

As for the all wave equations with a definite potential term in it, $C_0$ does not have any effect on waveform but the energy eigenvalues. This can be handled as a constant energy increase/decrease on all system regarding the choice of $C_0$. It already shows itself as an energy shift on Klein-Gordon equation in its most general form given in Eq.(9) as $q C_0$. This shows that if $q=0$ then there will be no electromagnetic interaction between the source and the particles created. Else, the interaction affects the energy eigenvalues by a multiplication by the constant of integration. So, it should be given in both the definition of electromagnetic four potential and wave equations. However, the choice of it does not directly affect either probability rate or particle density distributions that are the main results we would like to share. It has an indirect effect on energy as a shift and therefore should be considered on final energy eigenvalue calculations.

\section{Appendix-B}

Let us reconsider solutions of Eq. (25) under parametrizations of (34) that is the neutral limit.
In Eq. (34), the limits $(Q \rightarrow 0)$ and $(q \rightarrow 0)$ lead us to the same wave functions separately with real $L_Q$ and purely imaginary $E_Q$. This gives an insight that the produced particles from a neutral cosmic string may also be neutral. It is well-known that a real scalar field should be chargeless so one of the particle wave functions should be real if our assumption is true. This cannot be clearly seen in the current forms of solutions from (25) and (31). One can implement the relation equation between generalized Jacobi functions and hypergeometric functions:

\begin{equation}
    J_{\lambda'}^{(\alpha', \beta')} (x') = (2 sech(x'))^{1+\alpha' + \beta' - i \lambda'} \, _2F_1(\frac{1+\alpha' + \beta' - i \lambda'}{2},\frac{1+\alpha' - \beta' - i \lambda'}{2}; 1 - i \lambda'; sech^2(x'))
\end{equation}

Now let us take ($0<r<H^{-1}$)

\begin{equation}
 1-H^2r^2=sech^2(x'), \lambda'=-2iE'_Q, \alpha'=L'_Q, \beta'=-2\Tilde{M}'
\end{equation}

This choice gives us
\begin{equation}
    J_{-2iE'_Q}^{(L'_Q, -2\Tilde{M}')} (x') = 2^{\zeta} (1-H^2r^2)^{\frac{1+L'_Q-2\Tilde{M}'-2E'_Q}{2}} \, _2F_1(\zeta,\xi; 1 - 2E'_Q; 1-H^2r^2)
\end{equation}

where $\zeta, \xi$ are given in Eq. (26). So one of the neutral case solutions $\phi_3(r)$ in Eq.(30) can be rewritten as

\begin{equation}
\begin{matrix}
\phi_3(r) = N (-1)^{-E'_Q} (1-H^2r^2)^{-E'_Q} r^{\frac{-1}{2}+L'_Q} (1-H^2r^2)^{\frac{-1-L'_Q}{2}+\Tilde{M}'+E'_Q} J_{-2iE'_Q}^{(L'_Q, -2\Tilde{M}')} \\
= N (e^{i \pi})^{-E'_Q} r^{\frac{-1}{2}+L'_Q} (1-H^2r^2)^{\frac{-1-L'_Q}{2}+\Tilde{M}'} J_{-2iE'_Q}^{(L'_Q, -2\Tilde{M}')}(x')
\end{matrix}
\end{equation}

\noindent where $N$ is the normalization constant. Now since $E'_Q$ is purely imaginary in the neutral cosmic string case then terms $(e^{i \pi})^{-E'_Q}$ and $J_{-2iE'_Q}^{(L'_Q, -2\Tilde{M}')}(x')$ also the normalization constant and therefore finally $\phi_3(r)$ is purely real which makes it neutral necessarily. $\phi_3(r)$ can also be written as complex $\phi_1(r)$ and $\phi^*_1(r)$ mixed states.

Parity operator in this space is

\begin{equation}
    \hat{P} \Phi(t,r,\theta,\varphi) =  \hat{P} \Phi(t,r,\pi - \theta, \pi + \varphi)
\end{equation}

This keeps the radial wave equation same while affecting angular parts. Let’s give another way to show the solutions to the Eq. (7) in order to understand its characteristic under parity operation:

\begin{equation}
    \begin{matrix}
F_1(\theta) = csc(\theta) (-sin^2(\theta))^{\frac{\alpha+m}{2\alpha}} \, _2F_1(\frac{m-\alpha(1+l_\alpha)}{2 \alpha},\frac{m+\alpha(1+l_\alpha)}{2 \alpha};\frac{1}{2};cos^2(\theta)) \\
F_2(\theta) = cot(\theta) (-sin^2(\theta))^{\frac{\alpha+m}{2\alpha}} \, _2F_1(\frac{m-\alpha(-1+l_\alpha)}{2 \alpha},\frac{m+\alpha(2+l_\alpha)}{2 \alpha};\frac{3}{2};cos^2(\theta))
    \end{matrix}
\end{equation}
Hence
\begin{equation}
    \begin{matrix}
F_1(\theta-\pi) = F_1(\theta) \\
F_2(\theta-\pi) = - F_2(\theta)
    \end{matrix}
\end{equation}

Finally $e^{im(\pi + \varphi)} = (-1)^m e^{im\varphi}$. Since parity signature $\hat{P} \Phi = \pm \Phi$ depends on the choice of $m$, total wave function given in Eq. (6) can be either scalar or peseudoscalar.
Charge parity operator is given as

\begin{equation}
\hat{C} \Phi = \Bar{\Phi}
\end{equation}

where $\Bar{\Phi}$ denotes antiparticle. If $Qq > \frac{1}{2} + l_\alpha$ then Equation (15) shows that

\begin{equation}
\hat{C} \Phi_1 = \Bar{\Phi}_1 = \Phi_2
\end{equation}

Therefore, charge parity of the solutions $\Phi_1$ and $\Phi_2$ are +1.

These all calculations show that there is a potential for these solutions to correspond pseudoscalar mesons (e.g. pions) or similar (pseudo)scalar particle models.
This leads into two different potential observational consequences:
	\\1. Neutral Cosmic String\\
If the considered cosmic string exists in the nature and if it is neutral (the $Q\rightarrow0$ limit) then the given derivation of $\phi_3(r)$ shows that it is possible to observe chargeless scalar particles which may correspond to $\pi^0$ pion. In this case there is a potential of observing gamma rays which is decay result of neutral pions.
	\\2. Charged Cosmic String\\
If the considered cosmic string exists in the nature and if it is densely charged (the $Q \gg 1$ case) then there is a potential of complex particle wave solutions to occur that may correspond to observing electron and neutrino cosmic rays which occur due to decay of complex charged pions.

Showing how the results in this paper apply to different mass range broadens the scope of the work. If the created particles are the known mesons then their masses are expected  to lie within $M=100 MeV/c^2$ to $M=10 GeV/c^2$. This puts a lower limit on energy eigenvalues such that particle creation does not occur for the energy eigenvalues below its mass. In addition, lighter particles are more probable to be created.
In this study, Hubble constant ($H$) is taken to be position independent but it reduces by time. For example; current value, $H_0$, is assumed to be below $10^{-32} GeV$. Effect of particle mass on particle creation probability shows itself especially for a neutral CS in late-time de Sitter expansion (and before dark matter dominated era which is also regarded to have role in space expansion). Figure 6(a) shows how this probability changes with respect to varying $H$ for stationary particle pairs as the most expected outcome. Number of created particles are given in Figure 6(b) when string tension is $\mu =  1.35 \times 10^{21}$ (kg/m). Number of created particles depends on mass due to both probability, $P$ and product rate $\Gamma$.

\begin{figure}[H]
    \centering
    \includegraphics[width=0.8\textwidth]{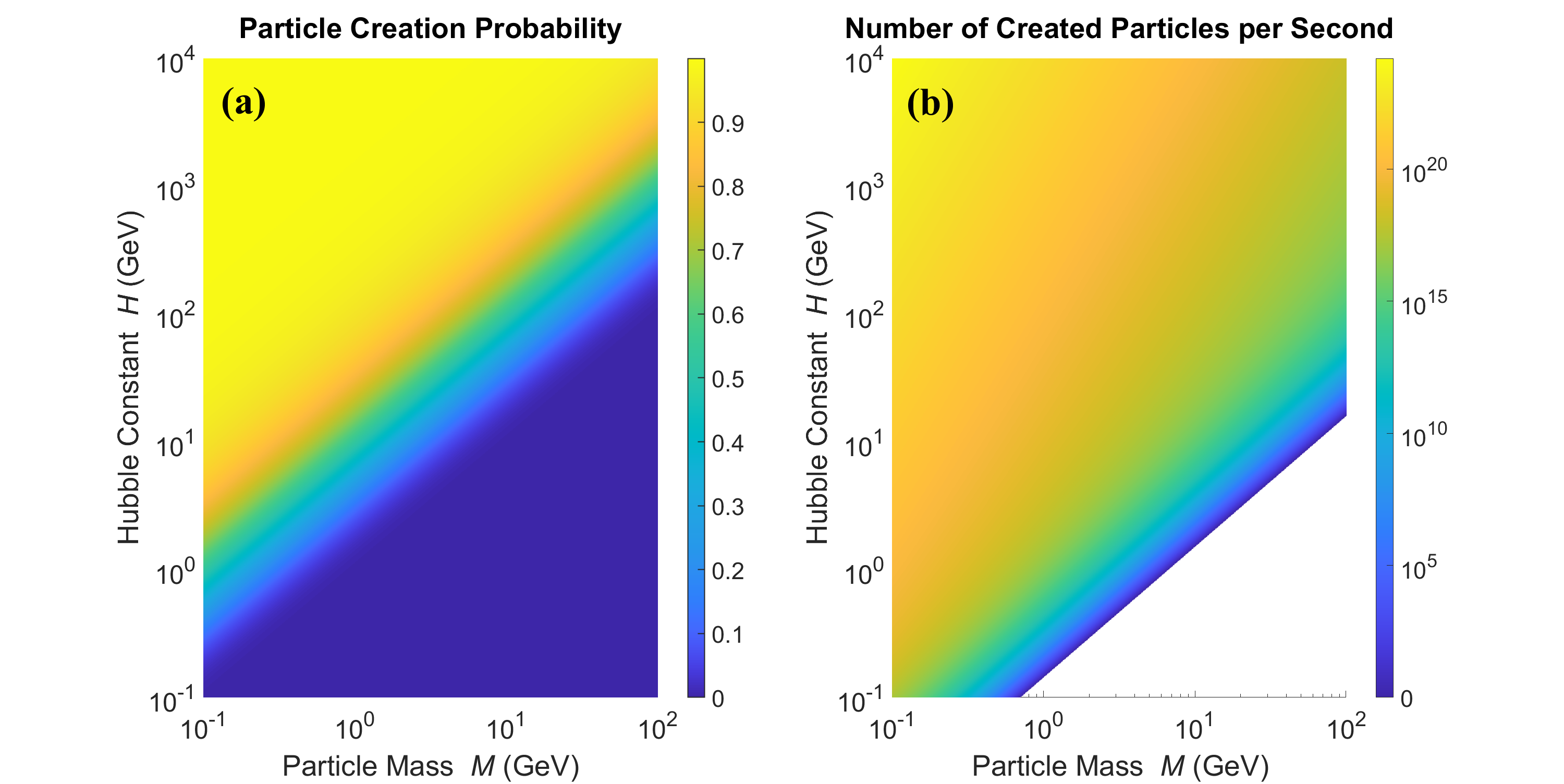} % Include the EPS file
    \caption{(a) Particle creation probability with respect to particle mass. (b) Expected value of number of created scalar particles $(N)$  within  a second. The white region shows where the particle creation vanishes. Figures are plotted for stationary particles.}
    \label{fig:6}
\end{figure}

\end{document}